\documentclass[oribibl]{svmult}

\usepackage{amssymb}

\usepackage{makeidx}         % allows index generation
\usepackage{graphicx}        % standard LaTeX graphics tool
\usepackage{multicol}        % used for the two-column index
\usepackage[bottom]{footmisc}% places footnotes at page bottom
\makeindex             % used for the subject index
\begin{document}

\title*{An Approach to Neutrino Radiative Transfer in Supernova Simulations}
\author{Christian Y. Cardall\inst{1,2}}
\institute{Physics Division, Oak Ridge National Laboratory, Oak Ridge,
 	TN 37831-6354
\texttt{cardallcy@ornl.gov} \and
Department of Physics and Astronomy, University of Tennessee,
	Knoxville, TN 37996-1200}
\maketitle

%Your text goes here. Separate text sections with the standard \LaTeX\
%sectioning commands.

\section{Core-collapse Supernovae: A Numerical Challenge}
\label{sec:1}
% Always give a unique label
% and use \ref{<label>} for cross-references
% and \cite{<label>} for bibliographic references
% use \sectionmark{}
% to alter or adjust the section heading in the running head

%Your text goes here. Use the \LaTeX\ automatism for your citations
%\cite{monograph}.

The approach to radiative transfer described in this contribution is being developed
for use in simulations of core-collapse supernovae. These events are the 
deaths of stars
more than about eight times as massive as the Sun, caused
by the catastrophic collapse of the star's core. This collapse is 
triggered in part by electron capture on heavy nuclei,
resulting in the emission of weakly-interacting particles called ``neutrinos.'' 
Early in the collapse
process these neutrinos escape freely, but eventually densities
are sufficiently high that even neutrinos are trapped. 
Collapse is finally halted when central densities reach a
few times the matter density of atomic nuclei. The newly-born neutron 
star---the compact object resulting from this process---is 
a hot thermal bath of dense nuclear matter, electrons, positrons,
neutrinos, and antineutrinos. Neutrinos and antineutrinos, 
having the weakest interactions among
the species present, are the most efficient means of cooling;
their emission accounts for virtually all of the gravitational
potential energy released during collapse.

Because neutrinos dominate the energetics of the supernova
process, neutrino radiative transfer is 
a central feature of the core collapse phenomenon. In particular,
energy transfer from neutrino radiation to infalling stellar matter
may be crucial to the supernova explosion mechanism \cite{colgate66,bethe85}. 
In addition to energy, neutrinos exchange a quantity called ``lepton number'' with the fluid; this composition variable affects the fluid's equation of state, and also has a strong influence on the relative abundances of nuclear species synthesized in
the supernova environment.

The radiative transfer of energy and lepton number is of particular interest in the semi-transparent region
near the neutron star surface. This is a region of transition from the
optically thick interior---where the diffusive neutrino field is nearly
isotropic---to the optically
thin exterior, where the neutrino radiation becomes strongly forward-peaked. 
Hence energy- and angle-dependent neutrino
transport is key to accurate modeling of core-collapse supernovae.

At least three groups have published reports of spherically
symmetric core-collapse supernova simulations with
energy- and angle-dependent neutrino transport.
Two of these groups---centered at 
the Max Planck Institute for Astrophysics 
\cite{rampp02} and
the University of Arizona 
\cite{thompson02}---employed a method 
in which angle-integrated and angle-dependent
neutrino transport equations are iterated to simulataneous
convergence.\footnote{See also H.T. Janka's contribution to this volume.} In contrast, the group centered at Oak Ridge National Laboratory
\cite{liebendoerfer02} 
performed a direct solution of the angle-dependent transport
equation, fully discretized in all variables in all terms.
All these efforts were grid-based (with the first two groups employing
``tangent rays'' to discretize the momentum angle).

The iteration of angle-integrated and angle-dependent 
transport equations---which might be called an
``iterated moment method''---works as follows. Zeroth and 
first angular moments of the neutrino transport equations
are formed, with energy dependence retained. 
The (energy-dependent) zeroth and first angular moments of 
the neutrino distribution thus become variables to be evolved.
The second and third angular moments also appear in these
equations; to close the system at the first moment,
the higher angular moments are expressed as numerical factors (the
so-called ``Eddington factors'') multiplying the zeroth moment.
The Eddington factors can be computed from the solution
of the angle-dependent transport equation; this is solved
with a simplified collision integral, which is expressed
in terms of the zeroth and first angular moments of the 
neutrino distribution. 
In summary: The moment equations need Eddington factors for
closure, which are obtained from the solution of the 
(simplified) angle-dependent transport equation; while the 
angle-dependent transport equation requires the zeroth and
first moments for its simplified collision integral. This
system is iterated to convergence.

The ``direct method'' of solving the transport equation
is not subject to a structural limitation of the iterated
moment method. In the direct method, all terms are 
discretized in all variables---time, space, energy, and
angles. In particular the angle dependencies
of neutrino scattering and pair production terms are fully represented,
while in the iterated moment method only the $l=0,1$
terms in a Legendre expansion of these 
collision kernels are employed; these are the only terms
in the expansion that can be constructed from the zeroth and
first moments of the neutrino distribution. (In principle,
the Eddington factors for the second and third
moments could be used to get two additional terms in the 
Legendre expansions, but this has not been implemented to date.) 
 In the
supernova environment, these truncated angular expansions
might not be accurate representations
of neutrino pair production and 
neutrino scattering by electrons and positrons---important 
processes 
in determining the spectra of neutrinos
emerging from the nascent neutron star. (Another 
approximation of undemonstrated safety---simplification or neglect of angular aberration in 
the angle-dependent
transport equation \cite{rampp02,thompson02}---seems less 
fundamental to the
iterated moment method itself.)

While energy- and angle-dependent neutrino transport is an
important advance even in spherical symmetry, 
the physics of stellar core collapse demands a spatially
multidimensional treatment. There is ample theoretical and
observational evidence for this conclusion (see for example
\cite{janka96}).

Inclusion of energy- and
angle-dependent neutrino transport in spatially multidimensional
simulations represents a significant
computational challenge. Consider the ``direct method''
mentioned above, in which the neutrino distribution functions
and all terms in the transport equation
are discretized in all variables. Assume azimuthal
symmetry; let the numbers of spatial zones in $(r,\theta)$ 
be $(256,128)$, and the numbers of momentum bins in energy and 
angle variables $(\epsilon,\vartheta,\varphi)$ 
be $(64,32,16)$. An implicit time evolution algorithm requires the 
solution of a large linear system, in which the matrix represents
the coupling between values of the neutrino distribution function
at different points on this five-dimensional grid.  
The spatial coupling of points having the same momenta is very sparse,
involving only nearest neighbors. But neutrino scattering and 
pair production terms in the collision integral involve dense
and extended coupling in momentum space; this means that the matrix contains
$256 \times 128$ dense blocks of size $(64 \times 32 \times 16)^2$. 
Storing all of these dense blocks in double precision would require 
a few $10^{14}$ bytes, beyond the capacity of today's terascale machines.
Moreover, the elements of the dense blocks are expensive to compute,
rendering unattractive an approach in which they are generated
``on the fly'' several times during each solution of the linear system.
These considerations motivate the algorithm presented
in this paper.

\section{Formalism for Radiative Transfer}
\label{sec:2}

Some details of a radiative transfer formalism and its finite differenced representation
will now be presented. 
Conservative formulations of radiative transfer---motivated by the importance
of accurately tracking energy and lepton number transfer in the supernova
environment---are discussed first. This is followed by a discussion of the finite differencing of the terms that do not depend on the velocity of the background fluid.

\subsection{Conservative Formulations of Relativistic Radiative Transfer}
\label{sec:3}

A radiative transfer calculation involves a variable describing the distribution in
phase space of the particles comprising the radiation.

The particle distributions we consider in some detail are 
the scalar distribution function $f$ and the
specific number density $\mathcal{N}$; we also mention the specific energy density
$\mathcal{T}$.
Each of these is taken to be a function of
spacetime coordinates $x^\mu$ and three-momentum variables
$u^{\hat{i}}$. (Greek and latin letters are spacetime and space indices
respectively. Hatted indices indicate quantities measured in an
orthonormal frame
comoving with the fluid with which the particle species interact, and
unadorned indices indicate components with respect to a global coordinate
basis.) The momentum variables $u^{\hat i}$ arise from a change of 
variables (e.g. to momentum space spherical coordinates) from $p^{\hat i}$,
the Cartesian spatial momentum components measured in a comoving orthonormal frame.

The scalar distribution function $f$ gives the the number of particles $dN$
in an invariant spacetime 3-volume element $dV$ and invariant
momentum space volume element $dP$ \cite{lindquist66}:
\begin{equation}
dN = f(x,{\bf p})\,(-v\cdot p)\,dV\,dP. \label{scalarDistribution}
\end{equation}
The quantities $x$, $v$, and $p$ are 4-vectors, and ${\bf p}$ is
the spatial 3-vector portion of $p$. The unit 4-vector $v$ is timelike, and defines the
orientation of $dV$:
\begin{equation}
dV = \sqrt{-g}\epsilon_{\mu\nu\rho\sigma}v^\mu\,d_1x^\nu\,d_2x^\rho\,
  d_3x^\sigma,
\end{equation}
where $g$ is the determinant of the metric tensor (taken to have
signature $-+++$) and $\epsilon_{\mu\nu\rho\sigma}$ is the Levi-Civita
alternating symbol ($\epsilon_{0123}=+1$). The momentum space volume element is
\begin{eqnarray}
dP &=& \sqrt{-g}\epsilon_{ijk} {\,d_1p^i\,d_2p^j\,
    d_3p^k \over (-p_0)} \nonumber \\
  &=& {1\over \epsilon}\left|\det\left(d{\bf p}\over d{\bf u}\right)\right|
    \,du^1\,du^2\,du^3, \label{momentumElement}
\end{eqnarray}
where specialization to momentum variables $u^{\hat i}$ in the comoving
frame has been made in the second line, and 
\begin{equation}
\epsilon \equiv p^{\hat 0} = \sqrt{|{\bf p}|^2 + m^2}
\end{equation}
is the particle
energy measured in the comoving frame ($m$ is the particle mass). 
The definition of $f$ in eq. (\ref{scalarDistribution}) makes
a convenient connection to nonrelativistic definitions of the
distribution function; for an equivalent but more geometric approach
see Ref. \cite{ehlers71}. The transport equation for electrically
neutral particles satisfied by $f$
is the Boltzmann equation 
\cite{lindquist66,ehlers71,mezzacappa89,cardall02}:
\begin{equation}
p^{\hat\mu}{\mathcal{L}^\mu}_{\hat\mu}{\partial f\over\partial x^\mu} -
{\Gamma^{\hat j}}_{\hat\mu\hat\nu}p^{\hat\mu}p^{\hat\nu}
{\partial u^{\hat i}\over\partial p^{\hat j}}{\partial f\over\partial 
u^{\hat i}} = \mathbb{C}[f]. \label{boltzmann}
\end{equation}
In this expression, ${\mathcal{L}^\mu}_{\hat\mu}$ is the transformation
between the coordinate basis and the comoving orthonormal basis; it 
involves a transformation to an orthonormal ``lab frame'' basis followed
by a Lorentz boost to the orthonormal comoving frame. The connection coefficients
in the orthonormal comoving basis are
\begin{equation}
{\Gamma^{\hat\mu}}_{\hat\nu\hat\rho} = {\mathcal{L}^{\hat\mu}}_{\mu}{\mathcal{L}^\nu}_{\hat\nu}{\mathcal{L}^\rho}_{\hat\rho}\, {\Gamma^{\mu}}_{\nu\rho} +
{\mathcal{L}^{\hat\mu}}_{\mu}{\mathcal{L}^\rho}_{\hat\rho}
{\partial {\mathcal{L}^\mu}_{\hat\nu}\over\partial x^\rho}.
\label{connectionComoving}
\end{equation} 
Because the transformation ${\mathcal{L}^\mu}_{\hat\mu}$ contains a 
Lorentz boost, the term ${\partial {\mathcal{L}^\mu}_{\hat\rho} / 
\partial x^\rho}$ gives rise to Doppler shifts and angular aberrations 
associated with this transformation. The coordinate basis connection coefficients,
\begin{equation}
{\Gamma^\mu}_{\nu\rho} = {1\over 2}g^{\mu\sigma}\left({\partial g_{\sigma\nu}
\over\partial x^\rho} + {\partial g_{\sigma\rho}\over\partial x^\nu} - 
{\partial g_{\nu\rho}\over\partial x^\sigma}\right), 
\end{equation}
where $g_{\mu\nu}$ are the metric components, give rise to energy shifts
and angular aberrations associated with spacetime curvature and the use of
curvilinear coordinates. The right-hand side of (\ref{boltzmann}) is the invariant collision integral.
%Dividing the Boltzmann equation by $\epsilon$
%and defining
%\begin{equation}
%n^{\hat\mu}\equiv \left(1, {p^{\hat i}\over\epsilon}\right),
%\end{equation}
%eq. (\ref{boltzmann}) becomes
%\begin{equation}
%n^{\hat\mu}{\mathcal{L}^\mu}_{\hat\mu}{\partial f\over\partial x^\mu} -
%{\Gamma^{\hat j}}_{\hat\mu\hat\nu}{n^{\hat\mu}p^{\hat\nu}}
%{\partial u^{\hat i}\over\partial p^{\hat j}}{\partial f\over\partial 
%u^{\hat i}} = {1\over\epsilon}\,\mathbb{C}[f]. \label{boltzmann2}
%\end{equation}
%(Note that $n^{\hat\mu}$ is not a 4-vector; the covariance of the
%Boltzmann equation is no longer manifest in this form.)

The Boltzmann equation can be put in a number-conservative form
\cite{cardall02}, which motivates the definition of the specific
particle number density $\mathcal{N}$. 
Specifically, eq. (\ref{boltzmann})
can be rewritten as
\begin{eqnarray}
{1\over\sqrt{-g}}{\partial\over\partial x^\mu}\left(\sqrt{-g}\,
p^{\hat\mu}{\mathcal{L}^\mu}_{\hat\mu}\,f\right) + & & 
\nonumber\\
\epsilon
\left|\det\left(d{\bf p}\over d{\bf u}\right)\right|^{-1} 
{\partial\over\partial u^{\hat i}}\left(-
{\Gamma^{\hat j}}_{\hat\mu\hat\nu}{p^{\hat\mu}p^{\hat\nu}\over\epsilon}
\left|\det\left(d{\bf p}\over d{\bf u}\right)\right|
{\partial u^{\hat i}\over\partial p^{\hat j}}\, f\right) &=& \mathbb{C}[f].
\label{numberConservative}
\end{eqnarray}
This form is called ``conservative'' because the left hand side
is expressed as divergences in spacetime and momentum space, so that
volume integrals of these terms are transparently related to surface terms. 
In particular, it is obvious that the momentum space divergence 
vanishes upon integration over $dP$ (given by eq. (\ref{momentumElement})),
yielding the equation for particle number balance:
\begin{equation}
{1\over\sqrt{-g}}{\partial\over\partial x^\mu}\left(\sqrt{-g}\,N^\mu \right)
= \int\mathbb{C}[f]\,dP,\label{numberBalance}
\end{equation} 
where 
\begin{equation}
N^\mu = \int{\mathcal{L}^\mu}_{\hat\mu} p^{\hat\mu}\,f\,dP 
\label{numberVector} 
\end{equation}
are the
coordinate basis components of the particle number flux vector. 
This motivates the definition of the specific particle number density 
$\mathcal{N}$, given by
\begin{equation}
\mathcal{N} \equiv {p^{\hat\mu}{\mathcal{L}^0}_{\hat\mu} \over \epsilon }\,f.\label{specificDensity}
\end{equation}
From eqs. (\ref{numberVector}) and (\ref{specificDensity}), we see that 
the specific number density is the contribution of each {\em comoving frame}
momentum bin to the {\em lab frame} number density:
\begin{equation}
N^0 = \int\mathcal{N}\;\epsilon\, dP.
\end{equation}
In terms of $\mathcal{N}$,
eq. (\ref{numberConservative}) can be rewritten as
\begin{eqnarray}
{1\over\sqrt{-g}}{\partial\over\partial x^\mu}\left(\sqrt{-g}\,
n^{\mu}\mathcal{N}\right) + & &
\nonumber\\
\left|\det\left(d{\bf p}\over d{\bf u}\right)\right|^{-1}
{\partial\over\partial u^{\hat i}}\left(-
{\Gamma^{\hat j}}_{\hat\mu\hat\nu}{\mathcal{L}^{\hat\mu}}_\mu
{n^{\mu}p^{\hat\nu}}
\left|\det\left(d{\bf p}\over d{\bf u}\right)\right|
{\partial u^{\hat i}\over\partial p^{\hat j}}\, \mathcal{N}\right) &=&
{1\over\epsilon}\,\mathbb{C},%\left[\mathcal{N}\over p^{\hat\mu}{\mathcal{L}^0}_{\hat\mu}\right],
\label{numberConservative2}
\end{eqnarray}
where we have defined
\begin{equation}
n^\mu \equiv \left(1, {p^{\hat\mu}{\mathcal{L}^\mu}_{\hat\mu} \over
p^{\hat\nu}{\mathcal{L}^0}_{\hat\nu}}\right).
\end{equation}
Note that $n^{\mu}$ is not a 4-vector.

Another reformulation of the Boltzmann equation is an 
energy-conservative form
\cite{cardall02}, which motivates the definition of the specific
particle energy density $\mathcal{T}$. This reformulation is not detailed here; suffice it to say that its momentum integral gives a transparent connection to the divergence of the particle stress-energy tensor.

The number- and energy-conservative formulations---which respectively facilitate
computation of lepton number and energy transfer to essentially machine 
accuracy---might be used in a couple of different ways. Both formulations could be solved separately, in order to nail down the transfer of both energy and lepton number. The values of the scalar distribution function implied by these two different distributions would then serve as a consistency check. Alternatively, only one of the formulations might be solved, with the analytic relationship between the two conservative formulations \cite{cardall02} being used to design finite difference expressions that provide consistency with the other formulation. This latter 
philosophy has been employed in the work of the group centered at Oak Ridge
\cite{liebendoerfer02}.

\subsection{Finite differencing of selected terms}
\label{sec:4}

Consider (\ref{numberConservative2}) in two
spatial dimensions in spherical coordinates, with the assumption of a static
(zero velocity) background and flat spacetime:
\begin{eqnarray}
{\partial{\cal N}\over \partial t} + {\cos\vartheta\over r^2}{\partial\over\partial r}
\left(r^2  \, {\cal N}\right) + {\sin\vartheta\cos\varphi \over r \sin\theta}{\partial\over\partial\theta}
\left(\sin\theta \, {\cal N}\right) -& & 
\nonumber \\
{1\over r \sin\vartheta}{\partial\over\partial \vartheta}\left(\sin^2\vartheta\, {\cal N}\right)
- {\cot\theta\over r}{\partial\over\partial\varphi}
\left(\sin\vartheta\sin\varphi \, {\cal N} \right) & &= {1\over\epsilon}\,\mathbb{C}.
\label{example}
\end{eqnarray}
In this equation, $(r,\theta)$ are the spatial radius and polar angle, $(\vartheta,\varphi)$ are momentum space angles, and $\epsilon$ is the particle energy.

There are two things to keep in mind in constructing a finite-differenced representation of (\ref{example}). First, one can take advantage of the conservative form to make numerical ``volume integrals'' transparently related to numerical ``surface integrals.'' Second, notice that for ${\cal N}$ spatially homogeneous,
the second and fourth terms of (\ref{example}) cancel, as do the third and
fifth terms. The finite difference representation should respect this cancellation.

A conservative differencing of the spatial divergence in (\ref{example}) is
\begin{eqnarray}
{1\over V_{i',j'}}(\cos\vartheta)_{\beta'}\left[(A_r\,{\cal N})_{i+1,j'} -
(A_r\,{\cal N})_{i,j'} \right] &+&
\nonumber \\
{1\over V_{i',j'}}(\sin\vartheta)_{\beta'}(\cos\varphi)_{\gamma'}\left[(A_\theta\,{\cal N})_{i',j+1} -
(A_\theta\,{\cal N})_{i',j} \right]&,& \label{spaceDivergence}
\end{eqnarray}
where the geometric factors are
\begin{eqnarray}
V_{i',j'}&=& 2\pi (r_{i'})^2 \sin(\theta_{j'}) (\Delta r)_{i'}  (\Delta\theta)_{j'}, \\
(A_r)_{i,j'} &=& 2\pi (r_i)^2 \sin(\theta_{j'})(\Delta\theta)_{j'}, \\
(A_\theta)_{i',j} &=& 2\pi r_{i'} \sin(\theta_j) (\Delta r)_{i'} .
\end{eqnarray}
In the finite-difference expressions in this subsection, the subscripts $(i,j)$ index spatial zones in the 
$(r,\theta)$ spatial coordinate directions, and subscripts $(\alpha,\beta,\gamma)$ index
momentum bins in the $(\epsilon,\vartheta,\varphi)$ momentum space coordinate directions.
Unprimed indices denote values evaluated on the surfaces (``edges'') of spatial
zones or momentum bins. The values of the coordinates $(r,\theta,\vartheta,\varphi)$ on the zone and bin edges are 
prescribed by the user. Primed indices denote values evaluated at zone or bin
centers. Given the ``edge values,'' the coordinates of the zone centers are taken to be
\begin{eqnarray}
r_{i'} &=& {1\over 2}\left(r_i^3 + r_{i+1}^3\right)^{1/3}, \label{rcenter}\\
\theta_{j'}&=& \arccos\left[{1\over 2}(\cos(\theta_j) + \cos(\theta_{j+1}))\right],
\label{thetacenter}
\end{eqnarray}  
and the zone widths are $(\Delta r)_{i'} = r_{i+1} - r_i$ and
$(\Delta \theta)_{j'} = \theta_{j+1} - \theta_j$.
%\begin{eqnarray}
%(\Delta r)_{i'} &=& r_{i+1} - r_i, \label{rwidth}\\
%(\Delta \theta)_{j'}&=& \theta_{j+1} - \theta_j. \label{thetawidth}
%\end{eqnarray}  
The definitions of $(\cos\vartheta)_{\beta'}$, $(\sin\vartheta)_{\beta'}$, and
$(\cos\varphi)_{\gamma'}$ in (\ref{spaceDivergence}) will be given below.
Finally, the values of ${\cal N}$ on zone surfaces in (\ref{spaceDivergence}) are
given by a particular linear interpolation of zone center values. 
This linear interpolation---which depends on the neutrino mean free paths---has the effect of shifting from ``diamond'' differencing in diffusive regimes to ``upwind'' (or ``donor-cell'') differencing in free-streaming regions; see \cite{liebendoerfer02} for details. 

A conservative differencing of the momentum space divergence in (\ref{example}) is
\begin{eqnarray}
-{1\over {\cal V}_{\alpha',\beta',\gamma'}}\left(1\over r\right)_{i'} \left[({\cal A}_\vartheta\,{\cal N})_{\beta+1,\gamma'} -
({\cal A}_\vartheta\,{\cal N})_{\beta,\gamma'} \right] &-&
\nonumber \\
{1\over  {\cal V}_{\alpha',\beta',\gamma'}}{(\cot\theta)_{j'} \over r_{i'}}
\left[({\cal A}_\varphi\,{\cal N})_{\beta',\gamma+1} -
({\cal A}_\varphi\,{\cal N})_{\beta',\gamma} \right]&,& \label{momentumDivergence}
\end{eqnarray}
where the momentum space ``geometric'' factors are
\begin{eqnarray}
{\cal V}_{\alpha',\beta',\gamma'}&=& (\epsilon_{\alpha'})^2 \sin(\vartheta_{\beta'}) (\Delta \epsilon)_{\alpha'}  (\Delta\vartheta)_{\beta'} (\Delta\varphi)_{\gamma'}, \\
({\cal A}_\vartheta)_{\beta,\gamma'} &=& (\epsilon_{\alpha'})^2 [\sin(\vartheta_{\beta})]^2 (\Delta \epsilon)_{\alpha'} (\Delta\varphi)_{\gamma'}, \\
({\cal A}_\varphi)_{\beta',\gamma} &=& (\epsilon_{\alpha'})^2 \sin(\vartheta_{\beta'}) 
\sin(\varphi_{\gamma})(\Delta \epsilon)_{\alpha'}  (\Delta\vartheta)_{\beta'}.
\end{eqnarray}
The bin center values $\epsilon_{\alpha'}$ and $\vartheta_{\beta'}$ are given
just as $r_{i'}$ and $\theta_{j'}$ in (\ref{rcenter}) and (\ref{thetacenter}). Also,
the momentum bin widths are given by the difference of the bounding edge
values, just as 
%in (\ref{rwidth}) and (\ref{thetawidth}) 
in the case of the spatial zones.
The definitions of $(1/r)_{i'}$ and $(\cot\theta)_{j'}$ will be given below.
The values of ${\cal N}$ on momentum bin edges are given by ``upwind'' or ``donor-cell'' differencing (see \cite{liebendoerfer02}). 

Finally, we show the remaining definitions needed to ensure that the
second and fourth terms in (\ref{example}), as well as the third and fifth terms,
cancel for ${\cal N}$ spatially homogeneous. Given the differencings defined
above, this is accomplished with the
following definitions:
\begin{eqnarray}
(\sin\vartheta)_{\beta'} &=& \sin(\vartheta_{\beta'}), \\
(\cos\vartheta)_{\beta'} &=&  {\sin(\vartheta_{\beta+1})^2 - \sin(\vartheta_{\beta})^2
\over 2 \sin(\vartheta_{\beta'}) (\Delta\vartheta)_{\beta'}}, \\
(\cos\varphi)_{\gamma'} &=&  {\sin(\varphi_{\gamma+1}) - \sin(\varphi_{\gamma})
\over (\Delta\varphi)_{\gamma'}}, \\
\left(1\over r\right)_{i'} &=& {(r_{i+1})^2 - (r_i)^2 \over 2 (r_{i'})^2 (\Delta r)_{i'}}, \\
(\cot\theta)_{j'} &=& {\sin(\theta_{j+1}) - \sin(\theta_{j})
\over \sin(\theta_{j'}) (\Delta\theta)_{j'}}.
\end{eqnarray}

\section{Radiative transfer algorithm and distributed-memory implementation}
\label{sec:5}

In this section classes of operators involved in
radiation transport are identified, and a strategy for implementing
them on distributed-memory computer architectures is described.

The equations of neutrino radiative transfer are the transport equation for whatever radiation particle distribution function is used, together with equations that describe lepton number and energy transfer (the latter are given by appropriate momentum integrals of the transport equation). 
The terms in these equations correspond
to operators acting on the discretized distribution function and transfer quantities. 
This can be expressed as 
\begin{equation}
F[y] = 0,
\label{discretized}
\end{equation}
where $y$ denotes the set of unknowns at a given time step:
$N_{\rm species}\times N_{\rm space}\times N_{\rm momentum}$
unknown values of the distribution functions of $N_{\rm species}$ neutrino species in $N_{\rm space}$ spatial zones and $N_{\rm momentum}$ momentum bins,
and $2\times N_{\rm space}$ unknown values of  energy and
lepton number transferred to the fluid in the $N_{\rm space}$ spatial zones.
The total operator $F$ has various pieces:
\begin{equation}
F = T + S + M + C.
\end{equation}  
The time derivative operator
$T$ relates unknowns at fixed position
${\bf x}$ and momentum ${\bf u}$ at different 
times $t$. The space derivative operator $S$ is linear, and
connects nearest neighbors 
in ${\bf x}$ at fixed ${\bf u}$ and $t$; similarly, 
the momentum derivative operator
$M$ is linear and connects nearest neighbors in ${\bf u}$ 
at fixed ${\bf x}$ and $t$. 
(The 
operators $S$ and $M$ are divergences in the conservative formulations.)
In the case of astrophysical neutrino transport, the collision operator $C$ 
is nonlinear due to neutrino-neutrino interactions and phase space 
blocking associated with the Pauli exclusion principle; and 
because of scattering and pair production and annihilation processes,
it exhibits extensive, nonlocal coupling in ${\bf u}$ at fixed ${\bf x}$
and $t$.

The large disparity between hydrodynamic ($\sim 10^{-3}$~s) and neutrino 
interaction ($\sim 10^{-10}$~s) time scales 
in the collapsed core of the supernova environment calls for implicit
evolution of the transfer equations. This means that in a time step
in which the system is evolved from time $t^n$ to $t^{n+1}$, the operators  
$S$, $M$, and $C$ are evaluated at $t^{n+1}$. The 
discretized transfer equations are then
\begin{equation}
T(y^{n+1},y^n) + S(y^{n+1}) + M(y^{n+1}) + C(y^{n+1}) = 0,
\label{implicit}
\end{equation}
where we have used the notation $y^{n} = y(t^n)$
and suppressed the dependence on space and momentum variables.
The dependence of $T$ on the values of $y$ at only two different
times indicates that a method that is first order in time is being used
(specifically, backward Euler).

Because of the nonlinearity of the collision operator $C$, 
(\ref{implicit}) is a set of nonlinear algebraic equations
for the discretized values of the distribution functions and transfer variables;
this system is solved with a Newton-Raphson iteration procedure. 
Specifically, (\ref{discretized}) is linearized:
\begin{equation}
J \cdot \Delta y = -F, 
\label{linearized}
\end{equation}
where $J \equiv \partial F/\partial y$ is the Jacobian matrix. In this linearized equation,
$J$ and $-F$ are evaluated at a guess $(y^{n+1})_{\rm guess}$ for
the value of the distribution function at the new time $t^{n+1}$. The
solution $\Delta y$ of this linear system provides a new guess 
$(y^{n+1})_{\rm new\ guess} = (y^{n+1})_{\rm guess} + \Delta y$.
This procedure is iterated to convergence of $y^{n+1}$ to the solution
of (\ref{implicit}).

To solve the linear problem at the heart of each Newton-Raphson
iteration, a simple fixed-point method is employed.
The basic idea is as follows.
Start with a guess
$(\Delta y)_{\rm guess}$ for the solution of (\ref{linearized}),
and compute the
residual $r$,
\begin{equation}
r = (-F) - J\cdot (\Delta y)_{\rm guess}.
\label{residual}
\end{equation}
Given $(J^{-1})_{\rm approx}$,
an approximate inverse of $J$, compute the correction $c$,
\begin{equation}
c = (J^{-1})_{\rm approx}\cdot r.
\label{correction} 
\end{equation}
Why the name
``correction''? Note that if one computes $c_{\rm exact}$ with
the exact inverse $J^{-1}$, (\ref{correction}), (\ref{residual}),
and (\ref{linearized}) give  
\begin{eqnarray}
c_{\rm exact}& =& J^{-1}\cdot(-F) - J^{-1}\cdot J\cdot (\Delta y)_{\rm guess}
\nonumber\\
& =& \Delta y - (\Delta y)_{\rm guess},
\end{eqnarray}
the difference between the exact solution $\Delta y$ 
to eq. (\ref{linearized}) and $(\Delta y)_{\rm guess}$.
When $c$ is computed with an approximate inverse it does not provide
the exact difference between $\Delta y$ and $(\Delta y)_{\rm guess}$,
but it does provide a new (and hopefully improved) guess 
$(\Delta y)_{\rm new\ guess} = (\Delta y)_{\rm guess} + c$.
This procedure is iterated to convergence of $\Delta y$ to the solution
of (\ref{linearized}).

Details of the implementation of this iterative fixed-point method for
the solution of the large linear system will now be given. In practice,
the inverses of two different approximate Jacobian matrices are applied in succession in each iteration (this also requires the computation of two residuals in each iteration). The first approximate Jacobian is 
\begin{equation}
J_{\rm momentum} = J_T + J_M + J_C, 
\end{equation}
which consists of the contributions to the Jacobian from the operators $T$, $M$, and $C$ in (\ref{implicit}). As previously described, this combination of operators densely couples different momenta ${\bf u}$ at fixed spatial position ${\bf x}$; hence $J_{\rm momentum}$ consists of $N_{\rm space}$ independent dense blocks---one for each spatial zone---and $\left(J_{\rm momentum}\right)^{-1}$ consists of individual inverses of these dense blocks. With the spatial grid partitioned among the many processors of a distributed-memory computer, the inversion of these separate blocks is trivially parallelized. The second approximate Jacobian is 
\begin{equation}
J_{\rm space} = J_T + J_S, 
\end{equation}
arising from the operators $T$, $S$ in (\ref{implicit}). By reasoning similar to above, $J_{\rm space}$ can be conceptualized as $N_{\rm momentum}$ independent matrices, but this time with sparse coupling, because the derivatives in $S$ only require nearest neighbors in space. Having chosen to partition the spatial grid, parallel solution of these independent ``spatial matrices'' requires an ``all-to-all'' communication to give each processor all the spatial data for its share of momentum bins; but the fact the matrices are sparse makes this communication manageable.

In addition to its simple structure, this fixed-point method has an important practical advantage over other iterative linear solver algorithms. As mentioned at the end of the first section, in spatially multidimensional problems simultaneous storage of all the dense blocks is impractical. The fixed-point algorithm outlined above can be structured so that each processor can construct a few dense blocks at a time, use them in all steps required in a given iteration, and discard them. In contrast, other linear solver algorithms seem to require dense blocks to be discarded and rebuilt multiple times in each iteration.

A code that implements the algorithm described above---written in Fortran 90, and using the MPI library for message passing---is being developed and tested at Oak Ridge National Laboratory, for eventual use in core-collapse supernova simulations. The implementation has been tested in both one and two spatial dimensions on a ``homogeneous sphere'' problem which has an analytic solution, with good results. The test has also been generalized to an ``inhomogeneous sphere'' problem, with emissivity and opacity varying in spatial polar angle. With regard to performance, it is found that inversion of the dense blocks dominates the computation; communication costs are not excessive. Because dense matrix solvers (e.g. the LAPACK library) are typically highly optimized, the dominance of the computation by dense blocks ensures that computational resources are used efficiently.

%
% BibTeX users please use
% \bibliographystyle{springer}
% \bibliography{bibliography_springer}
%
% Non-BibTeX users please follow the syntax
% the syntax of "referenc.tex" for your own citations
%\input{referenc}

%%%%%%%%%%%%%%%%%%%%%%%%%%%%%%%%%%%%%%%%%%%%%%%%%%%%%%%%%%%%%%%%%%%%%%

%%%%%%%%%%%%%%%%%%%%%%%%%%%%%%%%%%%%%%%%%%%%%%%%%%%%%%%%%%%%%%%%%%%%%%

\printindex
\end{document}